# Electron Beam Nanosculpting of Suspended Graphene Sheets

Michael D. Fischbein and Marija Drndić

*Department of Physics and Astronomy, University of Pennsylvania, Philadelphia, PA 19104*

We demonstrate high-resolution modification of suspended multi-layer graphene sheets by controlled exposure to the focused electron beam of a transmission electron microscope. We show that this technique can be used to realize, on timescales of a few seconds, a variety of features, including nanometer-scale pores, slits, and gaps that are stable and do not evolve over time. Despite the extreme thinness of the suspended graphene sheets, extensive removal of material to produce the desired feature geometries is found to not introduce long-range distortion of the suspended sheet structure.

Graphene, a two-dimensional carbon crystal, has been the focus of intense research since techniques were developed to extract it from graphite in the form of multi-layers[1] and single layers.[2] Graphene-based devices measured on substrates have revealed an impressive set of exotic electronic and optical properties with promising applications.[3-7] Furthermore, suspended graphene has been shown to have exceptionally high electron mobilities [8] and high strength.[9,10] Due to its single-atom-thickness and the relatively low atomic number of carbon, suspended graphene is emerging as powerful platform for transmission electron microscopy (TEM).[10,11,12] In addition to serving as a near-ideal substrate for TEM analysis,[13] it has been shown that electron-beam-induced-deposition (EBID) of carbon onto graphene can be achieved with high accuracy in a TEM.[14]

In this Letter we show that suspended multi-layer graphene sheets can be controllably nanosculpted with few-nanometer precision by ablation via focused electron-beam irradiation in a TEM at room temperature. We demonstrate nanopores, nanobridges and nanogaps. These examples and other nanometer-scale patterns of arbitrary design may prove useful in graphene-based electronic and mechanical applications. For instance, fabricating narrow constrictions in graphene layers is of interest for electronic property engineering.[15-23] Structures made by electron-beam irradiation are stable and do not evolve over time. Furthermore, we find that extensive removal of carbon does not introduce significant long-range distortions of the graphene sheet. Specifically, the sheets do not begin to fold, wrinkle, curl, or warp out of the focal plane during cutting.

Graphene sheets were extracted from graphite by mechanical exfoliation [2] on ~ 300 nm $SiO_2$ substrates coated with ~ 100 nm of PMMA and then transferred to a suspended ~ 50 nm-thick suspended $SiN_x$ membrane substrate.[24] Prior to transfer, arrays of ~ 1 μm square holes were patterned into the $SiN_x$ membranes by exposing the surface to a $SF_6$ reactive ion etch through a resist mask made by electron beam lithography. In order to transfer graphene sheets onto the $SiN_x$ membranes, we followed a method used by Meyer *et al.* [14] for transferring graphene to TEM-compatible holey carbon grids. After locating graphene sheets on the PMMA surface with an optical microscope, a drop of isopropanol is added to the surface. A $SiN_x$ membrane substrate is then placed onto the drop over a region containing graphene sheets, with its surface facing the PMMA surface. As the isopropanol evaporates, its surface tension brings the two surfaces into a close contact, which is further improved by heating at ~ 200 C for ~ 5 minutes. Finally, the PMMA is dissolved in acetone, which releases the graphene sheets on the PMMA side and allows them to transfer and stick to the $SiN_x$ membrane substrate.

Graphene sheets suspended over a hole in the $SiN_x$ membrane were identified in a TEM (JEOL 2010F operating at 200 kV). The number of graphene layers in a sheet could often be determined by imaging the edge of a folded region,[11] in a manner similar to counting the number of tubes in a multi-walled nanotube. We have worked with samples ranging in thickness roughly from 1 - 20 graphene layers, though the majority of graphene sheets used in this work were composed of ~ 5 layers. Using a method described previously, arbitrary patterns were created in the graphene sheets by increasing the TEM magnification to ~ 800,000x, condensing the imaging electron beam to its minimum diameter, ~ 1 nm, and moving the beam position with the condenser deflectors.[25] To avoid EBID of carbon, likely to occur for a spot-mode beam setting, nanosculpting was performed with the beam at cross-over in a diffusive mode. With the beam at cross-over, the current density measured on the imaging screen was ~ 50 pA/cm$^2$ which, after accounting for magnification, corresponds to an estimated ~ 0.3 pA/nm$^2$ at the sample position. The exposure of the graphene sheets to the beam was ~ 1 s/nm$^2$. All of the structures shown were made at room temperature.

Figures 1(a)-(c) show TEM images of a graphene sheet before and after creating a ~ 3.5 nm diameter nanopore by irradiating this spot with the condensed electron beam for ~ 5 s. We have also observed that very brief (~ 500 ms) exposure of graphene sheets to the condensed electron beam can be used to create a partial nanopore by removing a fraction of the graphene layers, while leaving other layers intact. A single nanopore is the simplest structure that can be made by ablation, yet nanopores have proven extremely valuable in studies of molecular translocation, DNA in particular.[26] Given that graphene is the thinnest possible membrane yet at the same time structurally robust [9] and impermeable,[27] nanopores in graphene sheets may be useful for achieving significant resolution enhancement in molecular translocation measurements. As shown in Fig. 1(d), multiple nanopores can be made in close proximity to each other, indicating that large arrays of closely packed nanopore arrays can be achieved. Since the average irradiation exposure time per nanopore is on the order of seconds, serial processing is not prohibitively time intensive and large arrays or more complicated geometries can be made quickly. Moreover, parallel fabrication with multiple electron beams would allow for substantial scalability.

All of the nanopores that we have made have a concentric ring-like structure extending several nanometers away from their edges. This ring-like structure, evident in Figs. 1(c) and 1(d), bares a close resemblance to the dark lines often observed at the edge of a folded graphene sheet, an example of which is shown in Fig. 1(e). The orientation of a folded graphene layer's edge is locally parallel to the TEM beam and consequently each layer in a folded graphene sheet introduces a dark line along the edge of the fold,[11] similar to what is seen at the radial edges of a multi-walled carbon nanotube. Intensity cross-sections [Figs. 1(f) and 1(g)] obtained from the images of the folded graphene sheet [Fig. 1(c)] and nanopore [Fig. 1(d)] reveal an average spacing between dark lines of 0.38 ± 0.02 nm and 0.39 ± 0.02 nm, respectively. These values are equivalent within the error introduced by finite TEM resolution and are close to the inter-layer distance of HOPG (~ 0.34 nm). These observations suggest that irradiation can induce coordinated inter-layer bonding between freshly exposed layer edges, leading in this case to an "inverted-onion"-like structure. Irradiation of carbon systems has been previously shown to be capable of inducing a variety of structural changes [28] and our results demonstrate that graphene sheets can provide a valuable initial system for deriving carbon morphologies.

Figure 2(a) shows two parallel ~ 6 nm wide lines, *i.e.*, regions where graphene has been removed, separated by ~ 25 nm. Starting with these lines, additional focused irradiation was used to gradually increase the lines' widths until their separation was reduced to ~ 5 nm, resulting in a "nanobridge" [Figs. 2(b) and 2(c)]. Although the final nanobridge is highly crystalline [Fig. 2(c)], the extensive exposure to irradiation may have induced significant inter-layer rebonding and atomic restructuring within individual layers. Nanobridges can be cut with the TEM beam to create a gap [Fig. 2(d)] with initial size less than a nanopore diameter but quickly increasing with continued irradiation. In the regions near the cut, irradiation induces morphological changes of the crystalline structure and, in particular, we observe that cut ends close completely, similar to fullerene capping observed for irradiated nanotubes.[28] Such carbon-based point contacts and nanobridges directly connected to a larger graphene structure may find use in mechanical and electrical applications.

In conclusion, we have demonstrated that suspended graphene sheets can be controllably nanosculpted with electron-beam irradiation. The ability to introduce features into suspended graphene sheets by electron-beam-induced cutting and reshaping with high spatial resolution expands their value as TEM compatible platforms and offers a route to fabricating graphitic structures for potential use in electrical, mechanical and molecular translocation studies.

**Acknowledgements**

This work has been partially supported by NSF (NSF Career Award DMR-0449533, NSF NSEC DMR-0425780 and MRSEC DMR05-20020), ONR YIP N000140410489, the Penn Genome Frontiers Institute and a grant with the Pennsylvania Department of Health. The Department of Health specifically disclaims responsibility for any analyses, interpretations, or conclusions.

**Figure Captions**

Figure 1

TEM images of a suspended graphene sheet is shown (a) before and (b) after a nanopore is made by electron beam ablation. (c) Higher magnification image of the nanopore. (d) Multiple nanopores made in close proximity to each other. (e) Folded edge of a graphene sheet showing lines corresponding to layer number. These lines are similar to those seen around the nanopores (Scale bars are 50, 50, 2, 10, 5 nm). (f) Average of intensity cross-sections taken along 6 different radial directions of the nanopore in (c), each starting at the edge and proceeding radially outward. (g) Average of 6 intensity cross-sections of the graphene sheet in (e), each taken perpendicular to and starting at the sheet edge.

Figure 2

(a) Two ~ 6 nm lines cut into a graphene sheet. (b) Electron irradiation is continued to create a ~ 5 nm wide bridge. (c) Higher resolution of the bridge shows clear atomic order. (d) Small gap opened in the nanobridge by additional electron irradiation. We note that the cut ends are closed. (Scale bars are 20, 10, 5, 5 nm).

Figure 1

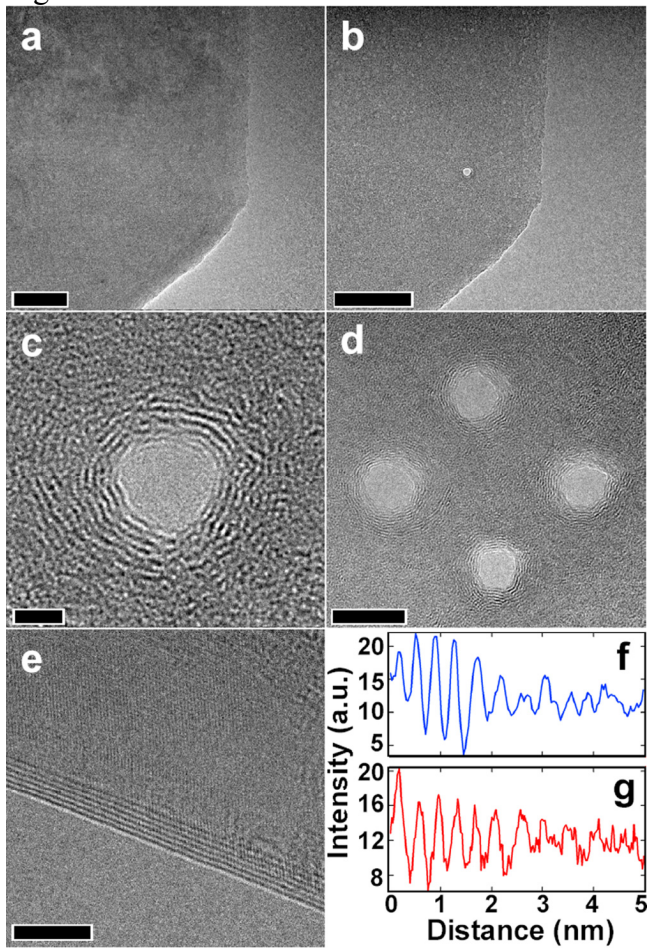

Figure 2

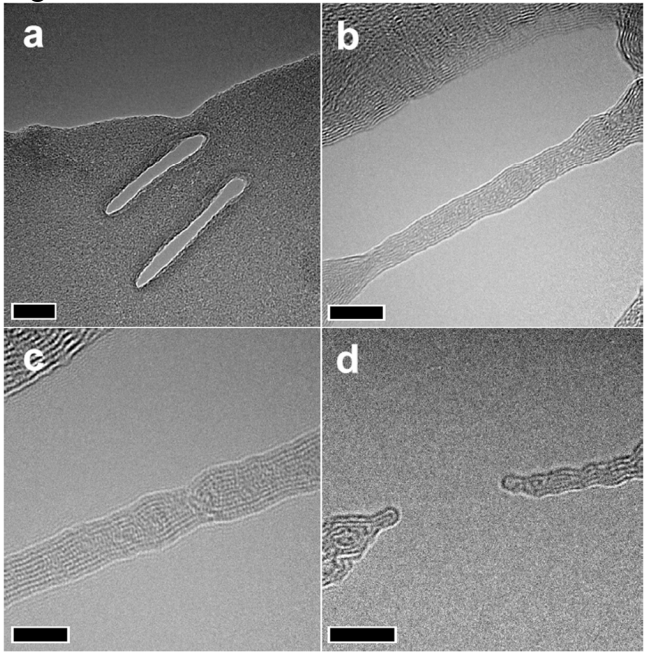